\documentclass[10pt, a4paper, twocolumn, showabstract]{naverlabseurope}

\usepackage{times}
\usepackage{latexsym}

\usepackage[T1]{fontenc}

\usepackage[utf8]{inputenc}

\usepackage{microtype}
\usepackage{enumitem}

\usepackage{ulem} 

\usepackage{inconsolata}

\usepackage{graphicx}

\usepackage{graphicx}
\usepackage{booktabs}
\usepackage{arydshln}
\usepackage{fontawesome5}
\usepackage{balance}
\usepackage{multirow}
\usepackage{makecell}

\usepackage{amsmath}

\title{Efficient Listwise Reranking with Compressed Document Representations}
\titlerunning{Efficient Listwise Reranking with Compressed Document Representations}

\correspondingauthor{herve.dejean@naverlabs.com}

\authors{ Hervé Déjean, Stéphane Clinchant }
\affiliations{NAVER LABS Europe}
\contributions{}
\website{https://github.com/naver/bergen}
\websiteref{\href{https://github.com/naver/bergen}}

\begin{abstract}

Reranking, the process of refining the output from a first-stage retriever, is often considered computationally expensive, especially when using Large Language Models (LLMs). A common approach to mitigate this cost involves utilizing smaller LLMs or controlling input length. Inspired by recent advances in document compression for retrieval-augmented generation (RAG),  we introduce RRK, an efficient and effective listwise reranker compressing documents into multi-token fixed-size embedding representations. 
Our simple training via distillation shows that this combination of rich compressed representations and listwise reranking yields a highly efficient and effective system. In particular, our 8B-parameter model runs 3$\times$–18$\times$ faster than smaller rerankers (0.6–4B parameters) while matching or outperforming them in effectiveness. 
The efficiency gains are even more striking on long-document benchmarks, where RRK widens its advantage further.

\end{abstract}

\begin{document}
\maketitle

\section{Introduction}

 Information Retrieval (IR) is typically understood as a two-part process: a \textit{first-stage} designed to swiftly locate pertinent documents for a specific query, followed by a more costly refinement phase called \textit{reranking}. 
Initially performed with cross-encoders \cite{gao_rethink_2021,nogueira2020passage}, LLMs are now used due to their strong performance. 

\begin{figure}[ht]
     \centering
     \includegraphics[width=\linewidth]{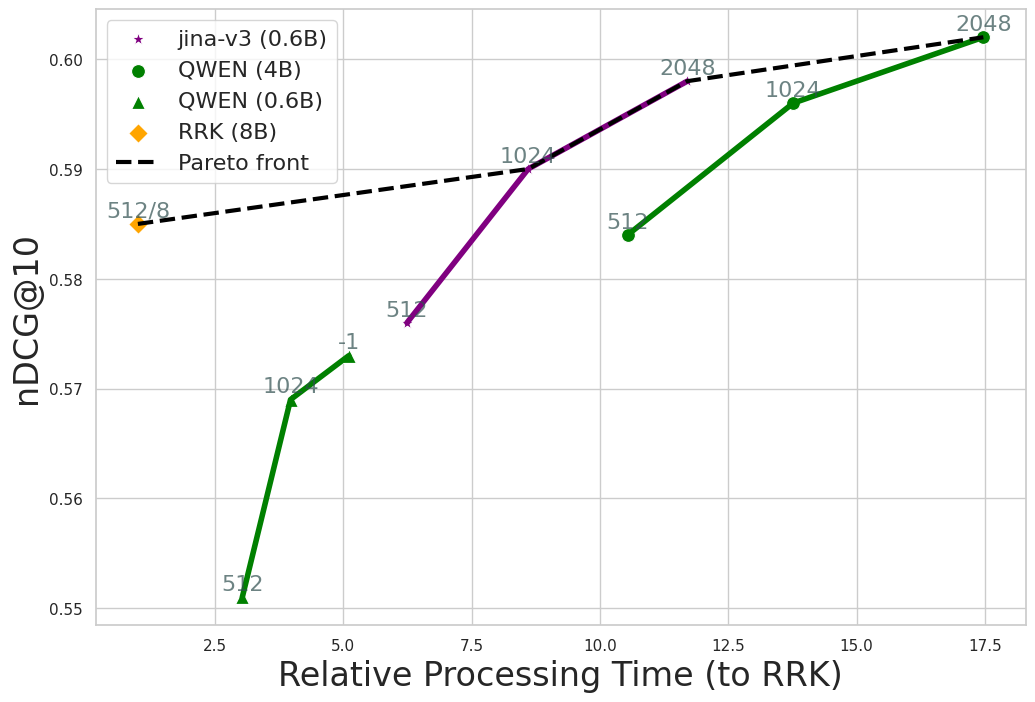}
     \caption{Efficiency/Effectiveness diagram for the BeIR collection. By integrating compressed document representation with listwise reranking, our 8B parameter RRK reranker outpaces all other rerankers in speed while maintaining robust effectiveness. Captions indicate document max-length. RRK compresses a 512-token document into an 8-token compressed version.}
     \label{fig:rrk}
\end{figure}

However, improving their efficiency is still an open challenge \cite{zhu2024largelanguagemodelsinformation} as 
LLM-based rerankers remain much less efficient than traditional cross-encoder rerankers \cite{zhuang_setwise_2023,déjean2024comp}.
Recent works have explored more efficient listwise reranking. A key step, introduced by \citet{gangi-reddy-etal-2024-first, zhuang_setwise_2023}, is to reduce latency by about 50\% by producing the full ranking in a single forward simply from the first-token logits. Other works trained smaller models and  \cite{wang2025jinarerankerv3lateinteractionlistwise} show that small rerankers trained on large-scale data can further improve efficiency or investigate document compression from first stage retriever embeddings  \cite{liu2025leveraging} and E2Rank \cite{E2rank2}.

In parallel, prompt compression methods have been proposed to accelerate LLMs in long-context settings, dialogue, and retrieval-augmented generation (RAG) \cite{icae,cocom,pisco,pilchen2025arcencoderlearningcompressedtext}. 
Such methods learn compact document representations that replace full retrieved texts with only a few tokens in the LLM prompt. Such findings naturally raise the question of whether such compressed representations could also support effective reranking.


\textbf{Contribution:} We introduce \textbf{RRK}, an efficient and effective listwise reranker based on compressed document representations. In contrast to prior approaches that rely on IR-based embeddings, RRK builds on the soft compression literature to produce rich multi-token document representations. RRK formulates listwise reranking directly over these compressed representations, drastically reducing the input length processed by the model and alleviating the efficiency bottleneck of LLM-based rerankers. Despite using an 8B-parameter backbone, RRK maintains strong effectiveness while achieving substantial efficiency gains. As illustrated in Figure~\ref{fig:rrk}, our model is \textbf{3×–18×} faster than state-of-the-art rerankers with substantially fewer parameters.

\section{Related Work}
\subsection{Efficient Rerankers}

First, LLMs showed strong potential as zero-shot rerankers:  RankGPT \cite{sun_is_2023}, built on GPT-4, achieved state-of-the-art performance as a zero-shot listwise reranker. 
\citet{qin_large_2023} show that listwise ranking with moderately sized open models often yields uninformative outputs, which motivates their pairwise reranking strategy combined with PRP-Sorting to improve both stability and efficiency. A common way to narrow this performance gap is through distillation: \citet{pradeep_rankzephyr_2023} fine-tune a Zephyr-7B model via knowledge transfer and obtain results comparable to GPT-4. More recently, \citet{zhuang_setwise_2023} systematically compare pointwise, pairwise, and listwise reranking, and propose a setwise prompting method that improves the effectiveness of zero-shot listwise approaches.


Another idea explored in \citet{liu2025matryoshkarerankerflexiblereranking} is to apply a Matryoshka architecture to rerankers, which allows to customize a reranker architecture by configuring the depth and width  of LLM,  achieving a 2× speed-up compared to the full model) with a sequence length of 1024. 


To make listwise LLM rerankers truly competitive, the original strategy of \citet{sun_is_2023}—explicitly \textit{generating} the identifiers of reranked documents—must be abandoned due to its high cost. While methods such as \citet{gangi-reddy-etal-2024-first} already reduce this cost by deriving the ranked order from the logits of the first generated token, more recent work \cite{E2rank2,wang2025jinarerankerv3lateinteractionlistwise} removes generation altogether. Instead, these approaches compute reranking scores directly from query and document representations.
E2Rank \cite{E2rank2} follows a similar direction. It trains document embeddings in two stages: first as retrieval embeddings, then with a joint first-stage retrieval and listwise reranking objective. At inference time, documents are replaced by embeddings, although the query representation still depends on the textual documents.

PE-Rank, proposed by \citet{liu2025leveraging} and most closely related to our work, improves listwise reranking efficiency by replacing textual documents with first-stage embeddings, following \citet{cheng2024xrag}. This results in a $4\times$–$6\times$ speedup, depending on document length, but still relies on a sequential generation step.

We compare both methods with our approach in Section~\ref{sec:perank}.


\subsection{Soft Compression}
In \cite{chevalier2023adapting}, the Autocompressor is introduced as a recursive context compression method trained on a language modeling objective. The method appends special compression tokens to the context and extracts their hidden states, enabling support for longer contexts and making it applicable to document compression in RAG-QA settings.

The In-Context Auto-Encoder (ICAE) \cite{icae} streamlines this idea by freezing the decoder, removing the recursive mechanism, and pretraining via a straightforward document auto-encoding task.

The xRAG method \cite{cheng2024xrag} reduces storage and computational overhead by reusing existing document embeddings from retrieval. Instead of learning new representations, it introduces a lightweight adapter that maps retrieval embeddings into the input space of a frozen decoder LLM.

Finally, PISCO \cite{pisco} introduces a more effective approach that relies entirely on knowledge distillation: both the compressor and decoder LLMs are trained to reproduce the outputs of a teacher model given raw text inputs. Remarkably, PISCO attains a 16× compression rate while maintaining high fidelity, with only a 0–3\% accuracy drop across a range of RAG-based question answering tasks. \cite{pilchen2025arcencoderlearningcompressedtext} shows that with a proper pretraining,  learning a compressor without fine-tuning or altering the target model's architecture  achieving state-of-the-art performance.

\section{Reranking Compressed Representation}

We build on the offline soft-compression framework PISCO \cite{pisco} to train a compressed reranking model called \textbf{RRK}\footnote{RRK: compressed version of \textbf{R}e\textbf{R}an\textbf{K}er}. RRK consists of two components: (i) a LoRA-finetuned PISCO compressor that maps documents to compressed token representations, and (ii) a LoRA-finetuned decoder reranker that assigns relevance scores to candidate documents conditioned on a query.

Let $\mathcal{D}=\{d_i\}_{i=1}^{N}$ denote a document collection and $q$ a query. The compressor $f_{\theta_c}: d_i \rightarrow \mathbf{c}_i$ maps each document to a sequence of compressed embeddings: $\mathbf{c}_i = (c_i^1,\ldots,c_i^l)$ where $l$ is the number of memory tokens. Given a query and a candidate set $D_k=\{d_1,\ldots,d_k\}$ retrieved by a first-stage retriever, the reranker $g_{\theta_r}(q,\mathbf{c}_1,\ldots,\mathbf{c}_k)$
produces relevance scores $s_i = g_{\theta_r}(q,\mathbf{c}_i)$
used to rank the documents.

\paragraph{Document Compression:}  After training the model, each collection is compressed \textbf{offline} using the finetuned  compressor: for each document $d_i$, a set of memory tokens $(m_1, \ldots, m_l)$ is appended, forming $(d_i; m_1, \ldots, m_l)$, which is fed to the compressor. The final $l$ hidden states of these memory tokens represent the document embeddings $\mathbf{c_i} = (c_i^s)_{s=1 \ldots l}$. In our case, we use a fix length of $l=8$ memory tokens, representing a x16 compression factor during PISCO training for a max-length of 128 tokens. We will show that the PISCO compressor naturally scales with longer documents (up to 2048 in our experiments, see Section~\ref{sec:longdoc}).

\paragraph{Listwise LLM Reranker} Our approach adopts an input representation similar to the Jina teacher \cite{wang2025jinarerankerv3lateinteractionlistwise}. The input sequence comprises query tokens (\textbf{in its textual representation}), followed by document memory tokens (\textbf{in their compressed representation}), and then the query tokens once more\footnote{Repeating the query at both the beginning and end may compensate for the lack of bidirectional attention and enhances the results}. 
For a query $q$ and candidate documents $\{d_1,\ldots,d_k\}$, the decoder input sequence is
\[
X =
(q ; \mathbf{c}_1 ; [\text{SEP}] ; \mathbf{c}_2 ; [\text{SEP}] ; \ldots ; \mathbf{c}_k ; [\text{SEP}]; q)
\]
Let $H = \text{Decoder}_{\theta_r}(X)$
denote the hidden states produced by the reranker. The query representation is extracted from the final token
$\mathbf{q} = H_{|X|}$
while the representation of document $d_i$ is taken from the hidden state $\mathbf{h}_i$ corresponding to the separator token following its compressed tokens. The ranking score is then computed by cosine similarity $s_i = \cos(\mathbf{q},\mathbf{h}_i).$

We use the RankNet listwise loss \cite{burges2005learning,gangi-reddy-etal-2024-first,E2Rank} to train RRK models.
Let $\mathcal{P}$ be the set of preference pairs $(d_i,d_j)$ where $d_i$ is preferred to $d_j$, the RankNet loss is parametrized by a temperature $\tau$\footnote{we use $\tau=1/8$} is:
\[
\mathcal{L}(q,D_k)
=
 \sum_{(i,j)\in\mathcal{P}}
\ \log(1+ \exp({ \frac{s_i - s_j}{\tau}}) )
\]

It is important to stress that the compressor is trained \textit{jointly} with the reranker, ie by backpropagating the ranking loss through the scores $s_i = g_{\theta_r}(q,f_{\theta_c}(d_i))$. After training, the documents are compressed offline and the decoder can be applied during inference to sort document by their scores.

Importantly, compression enables to reduce the reranking complexity. Let $k$ is the number of documents to be reranked and $|q|$ denote the number of query tokens and $|d|$ the average document length. A standard LLM reranker processes sequences of length $|q| + k|d|$, leading to an attention complexity of $\mathcal{O}((|q| + k|d|)^2)$. Our model instead processes sequences of length $2|q| + k(l+1)$, where $l$ is the number of memory tokens. The resulting complexity becomes $\mathcal{O}((2|q| + k(l+1))^2)$. Since $l \ll |d|$ (e.g., $l=8$ vs.\ $|d|\approx 200$ on average for most BeIR collections), this substantially reduces the quadratic attention cost of the reranker.



\section{Experimental Framework}
\label{sec:training}
\paragraph{Distillation and Teacher}
Training high-quality rerankers usually requires large labeled datasets and carefully selected negatives from multiple retrievers \cite{cao2024recentadvancestextembedding}, making comparisons difficult and reducing reproducibility. Instead, we adopt a simpler approach based on distillation from a state-of-the-art reranker. Our goal is to evaluate whether rerankers trained on compressed document representations remain effective while improving efficiency. To this end, we distill a teacher into several models trained on a small dataset, including a base LLM, a ModernBERT baseline, and RRK. This setup enables a fair comparison between compressed and textual representations while remaining simple and reproducible. 



In order to select our teacher,  we performed a set of evaluation using various first-stage and rerankers.
Based on those results, 
we choose the SPLADE-V3 \cite{lassance2024spladev3newbaselinessplade}, a fast model proven to be robust out-of-domain, and the jina reranker V3\footnote{jinaai/jina-reranker-v3}, a listwise reranker based on a Qwen 0.6B  backbone \cite{wang2025jinarerankerv3lateinteractionlistwise}, which performs as well as the Qwen3-4B reranker. \citet{qwen3embedding} shows that larger LLMs (8B) perform similarly in terms of effectiveness.


\paragraph{Training Set:}For our training collection, we first use the traditional MS MARCO (passage) dataset \cite{msmarco}. 
The training collection, which consists of a set of queries  and an appropriate document collection (without the need for relevance judgments), is processed using the selected first-stage retriever and reranker. For each query, we identify the top 50 documents produced by the reranker used as teacher. For the query set, we utilize the  0.5 million training queries, pairing each query with 16 documents randomly selected from the top 50 documents provided by the retriever. 
To evaluate this choice and to boost the results, a second collection, provided by \citet{E2rank2}, is also employed. This collection, derived from BGE-M3 \cite{qwen3embedding}, comprises 150,000 queries, each associated with 16 documents scored by Qwen-32B used as zero-shot teacher reranker. 

\paragraph{Qwen Backbone:} We select a Qwen backbone since the Qwen family provides multiple rerankers (our teacher and baselines). We train a PISCO model using the Qwen-2.5 8B-instruct model. The training of this PISCO-Qwen model precisely follows the methodology outlined by \citet{pisco}.

\paragraph{Pointwise LLM Reranker (RRK PW)} We also present a pointwise approach as comparison. At reranking stage, the compressed documents embeddings $\mathbf{c}$ are loaded and fed to the Decoder (Qwen-2.5 8B) finetuned for reranking. The model’s decoder takes as input the query (\textbf{in its textual representation}) along with the \textbf{compressed representation} of the document $\mathbf{c}$, and generates a score. We train the model with a mean squared error (MSE) loss to reproduce the scores of a teacher reranker. Specifically, a linear layer maps the final-layer representation of the last token to a scalar score.

To train our RRK models, we conduct training over 2 epochs, as additional epochs did not yield significant improvements. The finetuning takes 48h using 1 A100 GPU with 16 document per query, a batch size of 2, a learning rate of $1 \times 10^{-4}$, and gradient accumulation of 16.  For the pointwise version we use 4 documents per query and a batch size of 8 (similar training time).
Regarding latency, all computations were performed on a single A100 GPU unless otherwise specified. Across all configurations, we used a Across all configurations, we used a batch size of 128, corresponding to over 90\% GPU memory utilization, except for the baseline models at input length 512, for which we increased the batch size to 256.
At inference, RRK requires access to the compressed representations, which accounts for less than 10\% of total reranking time. We store embeddings in a Hugging Face dataset and fetch document embeddings with the \texttt{select()} function.

\section{Evaluation}
For evaluation, we use standard IR benchmarks: TREC Deep Learning 2019/2020 \cite{craswell2020overview,craswell2021overview} and BeIR \cite{Thakur2021BEIRAH}. For long-document ablations, we use the MS MARCO Document variants of TREC-DL 2019/2020. Following \cite{déjean2024comp, E2Rank}, we exclude the BeIR ArguAna collection, which targets counter-argument retrieval. We rerank the top-50 candidates from SPLADE-v3, which yields higher effectiveness than using the BM25 top-100 while also improving efficiency by reranking fewer documents. We report nDCG@10 ($\times 100$ for legibility) on all datasets, and measure efficiency using the \textit{latency ratio}, defined as the relative processing time with respect to RRK, which is always the fastest.
\textbf{Listwise rerankers} (RRK, Jina-V3) are mentioned with a $\mathbf{\dagger}$ in the Tables.

As baselines, we include the publicly available Qwen-3 0.6B and 4B pointwise rerankers \cite{qwen3embedding}. Since these models are trained on substantially larger datasets (7M labeled and 12M synthetic examples), our goal is primarily to compare latency rather than to exceed their effectiveness, while tolerating a possible effectiveness gap. We also compare the teacher model, the listwise Jina-v3 reranker, to the pointwise and listwise RRK in terms of effectiveness and efficiency.
This comparison is particularly informative because non-compressed listwise rerankers are typically slower than their pointwise counterparts. 

To isolate the impact of compression versus textual inputs, we also train two textual-input models under the same setting: Qwen2.5 8B (the backbone used for RRK) and ModernBERT-large \cite{warner2024smarterbetterfasterlonger}, a “smaller, better, faster, longer” \textit{sic} bidirectional encoder that is competitive in both effectiveness and efficiency. The Qwen2.5 8B textual model serves as an upper bound on effectiveness despite its impractical latency, while ModernBERT provides a strong encoder-only baseline for latency comparisons. Both models are trained with a pointwise loss, since a non-compressed listwise reranker is always slower than its pointwise counterpart.

\section{Results}
\begin{table}[h]
    \centering
    \resizebox{0.5\textwidth}{!}{
   
     
\begin{tabular}{lrrrr}
    \toprule
    \textbf{Model} &\textbf{Len.} & \textbf{nDCG@10} & \textbf{Ratio} & \textbf{s/q} $\downarrow$ \\
    \midrule
      \multicolumn{5}{l}{\textbf{RRK  Rerankers (QWEN2.5-8B)} }\\
    \midrule
        RRK  $\dagger$ & 512   & 58.4 & 1    & 0.06 \\
        RRK  & 512   & 57.5 & x3   & 0.21\\
    \midrule
     \multicolumn{5}{l}{\textbf{Public Rerankers} }\\
    \midrule 
    Jina-v3  $\dagger$ (QWEN3-0.6B) & 512     & 57.6 & x6   & 0.44 \\
         & 1024    & 59.0 & x8   & 0.53 \\
         & -1      & 59.8 & x11  & 0.72 \\       
    Qwen3-0.6B & 512 & 55.1 & x3   & 0.18 \\
            & 1024 & 56.9 & x4   & 0.24 \\
            & 2048 & 57.3 & x5   & 0.31 \\             
    Qwen3-4B & 512    & 58.4 & x10  & 0.64 \\
          & 1024   & 59.6 & x14  & 0.84 \\
          & 2048   & \textbf{60.2} & x17  & 1.0\\
\midrule
  \multicolumn{5}{l}{\textbf{Fine-tuned models without compression}} \\
\midrule
    ModernBert-Large & 512 & 57.2 & x2   & 0.13 \\
    Qwen2.5-8B & 512  & 59.7 & x20  & 1.26 \\
 
    \bottomrule
\end{tabular}
   
    }
        \label{tab:qwen}
         \caption{Evaluation and latency of our models for the BeIR collection (12 datasets).  nDCG@10 and latency ratio relative to RRK. While public models perform the best  when considering a long input length (2048), their latency compare badly to our RRK models using compressed document representation. $\dagger$: listwise reranker. See also Figure~\ref{fig:rrk}. }
           \label{tab:qwen}
\end{table}

The main results are presented in Table~\ref{tab:qwen} while Figure~\ref{fig:rrk} illustrates the efficiency-effectiveness trade-offs. 
We report the average processing time per query across the full BeIR dataset (31,828 queries):   
detailed BeIR results are given in Table \ref{tab:rerankerbeir}.

\begin{table*}[ht]
    \centering
    \resizebox{1.0\textwidth}{!}{
    \begin{tabular}{l|c|c|c||c|c||c|c}
    \toprule
    &
    \textbf{SPLADE-V3}
    &  \multicolumn{1}{c|}{\textbf{Jina v3$\dagger$}}
    & \multicolumn{1}{c||}{\textbf{Qwen3 4B}}
    &  \multicolumn{1}{c|}{\textbf{ModernBERT}}
    &  \multicolumn{1}{c||}{\textbf{Qwen-2.5 8B}}
    & \multicolumn{1}{c|}{\textbf{RRK}}
    & \multicolumn{1}{c}{\textbf{RRK$\dagger$}}
    \\
  &(retriever)&(teacher)&&(text)&(text)&&\\
    \midrule
    \textbf{TREC} & \\
    \midrule
    DL 19       & 72.3 & 75.3 & 76.5 & 76.3 & \textbf{77.9} & 77.5 &  75.8 \\
    DL 20       & 75.4 & 66.8 & 75.3 & 76.7 & \textbf{79.0} & 77.6 &  77.1 \\
    \midrule 
    \textbf{BeIR} &\\
    \midrule
    TREC-COVID  & 74.8 & 87.8 & 88.1 & 89.0 & 87.7 & 86.5 &  \textbf{89.3} \\
    NFCorpus    & 35.7 & 36.7 & 38.6 & 38.1 & \textbf{38.7}  &\textbf{38.7}  & 37.2 \\
    NQ          & 58.6 & {72.5} &\textbf{ 77.5} & 66.0 & 72.3 & 66.3 &  70.2 \\
    HotpotQA    & 69.2 & \textbf{80.3} & 79.1 & 75.4 & 78.4 & 73.7 &  76.3 \\
    FIQA        & 37.4 & 46.1 & 46.9 & 47.6 & \textbf{49.4} & 47.5 &  45.3 \\
    Touché 2020-v2 & 29.3 & 32.8 & 32.5 & \textbf{35.2} & 32.7 & 31.1  & 33.6 \\
    Quora       & 81.4 & \textbf{89.9} & 84.9 & 86.0 & 89.2 & 86.8 &  87.6 \\
    DBPedia     & 45.0 & 48.7 & 48.3 & \textbf{52.0 }& 48.7 & 49.9 &  {51.2} \\
    SCIDOCS     & 15.8 & {22.2} & \textbf{23.3} & 19.5 & 21.3 & 19.6  & 21.1 \\
    FEVER       & 79.6 & \textbf{91.6} & 90.5 & 88.4 & 89.4 & 84.8 &  85.3 \\
    Climate-FEVER & 23.3 & {33.7} &\textbf{ 39.5} & 25.3 & 28.3 & 27.0  & 28.1 \\
    SciFact     & 71.0 & 75.6 & 77.4 & 75.4 & \textbf{77.6} & 76.4 &  75.3 \\
    \midrule
    \textbf{AVG} & 51.8 & 59.8 & \textbf{60.2} & 57.9 & 59.4 & 57.5  & 58.4 \\    
    \bottomrule
    \end{tabular}
    }
    
    \caption{Detailled evaluation for the TREC DL 19/20 and  BeIR collections ($nDCG@10 * 100$). The top 50 of the SPLADE-v3 retrieved documents is used as input. For the public rerankers we report their best performances with an input length of 2048. Trained models' performances are reported with an input length of 512 (no improvement with longer input for those datasets). $\dagger$: listwise rerankers.
    }
    \label{tab:rerankerbeir}
\end{table*}

Our key result is that the listwise RRK model is the fastest system—over $2\times$ faster than ModernBERT and up to $17\times$ faster than 4B parameter models—while maintaining strong effectiveness; RRK outperforms all baselines using 512 tokens as input, and outperforms the 0.6B rerankers in both effectiveness and efficiency in nearly all input length configurations.
Figure~\ref{fig:rrk} clearly illustrates RRK’s speed advantage.

The fine-tuned models behave as expected: the Qwen2.5 8B textual model matches its teacher’s effectiveness, while ModernBERT-large is substantially faster ($4\times$–$6\times$) with acceptable effectiveness. 

The pointwise RRK variant attains a lower effectiveness to RRK and is less competitive overall: ModernBERT performs similarly but is twice faster, and Qwen-3 0.6B is comparable. Nonetheless, despite its 8B size, its latency is remarkable relative to the 0.6B models.

For publicly released rerankers, effectiveness is highly sensitive to input length ($-2$ points), as well as  latency  (about $2\times$ slower when increasing from 512 to 2048 tokens). This sensitivity  is surprising  on BeIR, where many sub-collections contain relatively short documents. We do not observe this phenomenon with our trained rerankers, which may be a consequence of using MS MARCO, a collection of short documents.  For this reason, we present RRK results with an input length of 512.

Overall, combining compressed representations with a listwise objective yields a highly efficient reranker without sacrificing effectiveness.

\subsection{Reranking Long Documents}
\label{sec:longdoc}

\begin{table*}[h]
    \centering
    \small
    \begin{tabular}{lllllccccccccc}
        \toprule
        \multirow{2}{*}{\textbf{Doc Length}} & \multicolumn{2}{c}{\textbf{RRK-MS$\dagger$}} & \multicolumn{2}{c}{\textbf{RRK$\dagger$}} & \multicolumn{2}{c}{\textbf{Jina-v3$\dagger$}} & \multicolumn{2}{c}{\textbf{Qwen3 4B}} & \multicolumn{2}{c}{\textbf{Qwen3 0.6B}} & \multicolumn{2}{c}{\textbf{MBerT}}\\
        &  nDCG & Lat & nDCG & Lat & nDCG & Lat & nDCG & Lat &nDCG & Lat&nDCG & Lat \\
        \midrule
        \multicolumn{11}{l}{\textbf{MS MARCO DOCUMENT DL19}} \\
        \midrule
        512  & 68.5 & 1&70.6 & 1&62.6 & x11 & 63.0 & x13 & 60.3 & x4 & 69.7 & x6 \\
        1024 & 68.6 & 1&\textbf{72.1} & 1&66.4 & x24 & 66.9 & x21 & 64.5 & x5 & 70.0 & x7 \\
        2048 & 68.6 & 1&72.0 & 1&68.3 & x58 & 70.0 & x37 & 65.5 & x10 & 69.7 & x8 \\
        \midrule
        \multicolumn{11}{l}{\textbf{MS MARCO DOCUMENT DL20}} \\
        \midrule
        512 & 66.1 & 1&67.7 & 1  &60.0 & x12 & 58.5 & x13 & 55.0 & x4 & 66.3 &x6 \\
        1024 & 66.5 &1 &67.0 &1  &62.6 & x24 & 62.8 & x20 & 59.4 & x6 & 66.4 & x7 \\
        2048 & 67.0 &1 &\textbf{68.6} &1  &64.5 & x59 & 66.9 & x35 & 63.7 & x10 & 67.4 & x8 \\
        \bottomrule
    \end{tabular}
   
     \caption{ Reranking long documents: RRK performs very well for the MS Marco Document DL 19/20 collections, and its efficiency is even more remarkable (See also Figure~\ref{fig:doc19}). nDCG@10, latency ratio relative to RRK.}
      \label{tab:longdoc}
\end{table*}

\begin{figure}[h]
    \centering
    \begin{minipage}{0.45\textwidth}
        \centering
        \includegraphics[width=\textwidth]{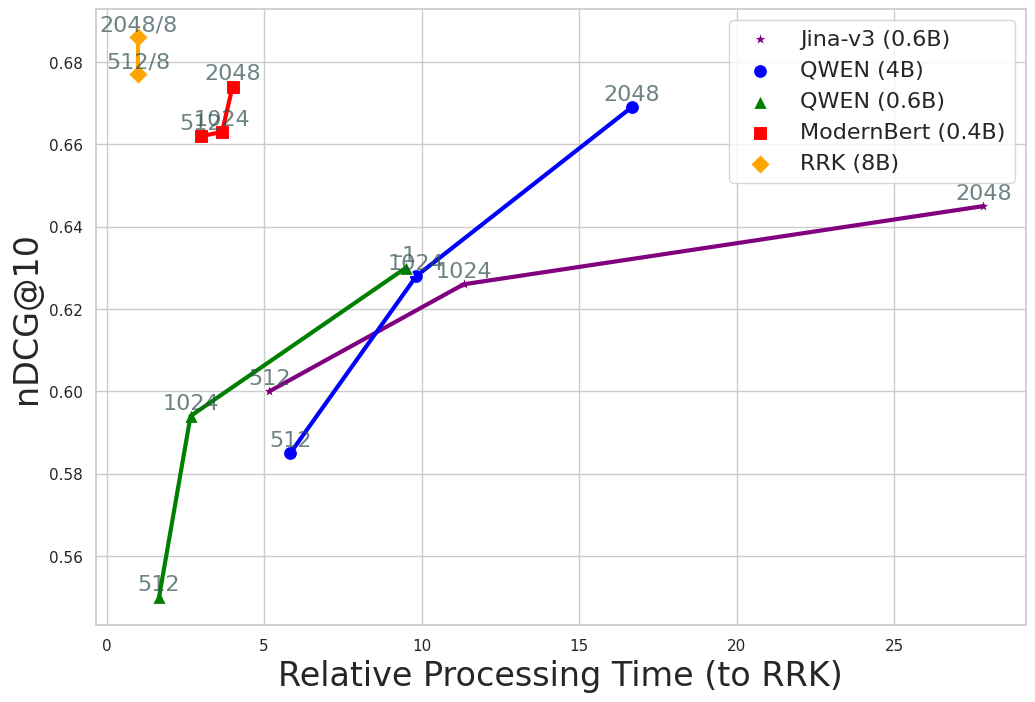} 
        \caption{Efficiency/Effectiveness diagram for the MS-MARCO Document TREC 2020. Captions indicate document max-length}
        \label{fig:doc19}
    \end{minipage}\hfil
\end{figure}

We now focus on datasets where compression significantly enhances efficiency: specifically, datasets containing "long" documents (those exceeding 512 tokens, a typical length for encoder-only rerankers). We employ the MS-MARCO document collection along with the TREC 2019/2020 evaluation set for this purpose. The average document length is approximately 1000 tokens \cite{ma2023finetuningllamamultistagetext}. We reranker the top 50 documents provided by Splade-v3 as before. 
In this experiment, we test several input lengths for all models: our baselines as well as RRK: 512, 1024 and 2048.
The results, presented in Table~\ref{tab:longdoc} and Figure~\ref{fig:doc19}, are surprising: the open rerankers underperform. 
Our RRK models, along with our trained ModernBert, demonstrate impressive effectiveness.
Interestingly, RRK, although trained to compress short documents (128 tokens in the PISCO setting), effectively handles long documents and responds favourably to increased document lengths within these collections, a phenomenon not observed with the BeIR collection. RRK is capable of  achieving a remarkable compression factor up to 256 (the best result for DL20 is obtained with an input length of 2048), being ten times faster than the pointwise Qwen 0.6B reranker in this setting.

Figure~\ref{fig:doc19} further shows that long documents substantially increase the latency of non-compressed listwise rerankers: Jina-v3, our 0.6B-parameter teacher, becomes as slow as the  pointwise Qwen 4B reranker. This confirms that listwise reranking, when used without compression, does not by itself provide sufficient efficiency benefits.

Our small  ModernBert model performs exceptionally well in this context, offering competitive performance. We anticipated better results from the public rerankers, particularly Jina-V3, which was meticulously trained to handle long documents. We suspect that the underwhelming performance of these public rerankers may be  due to the diverse training material utilized, whereas our dataset is domain-specific (MS MARCO). Note that these models also underperform for the passage TREC DL 19/20 collections (Table~\ref{tab:rerankerbeir}).

\subsection{Ablations}
We perform two types of ablations: the first examines the impact of the training dataset and teacher, while the second investigates the process of generating the PISCO compressor.
While working to enhance the effectiveness of our RRK, one evident strategy was to incorporate additional training data by utilizing the training set provided by \cite{E2Rank}. This dataset comprises 150,000 queries and uses a different teacher model (Qwen3-32B). On its own, it yields worse results compared to our MS-MARCO dataset, even worst than a MS-MARCO subsample of the same size. We hypothesize that E2Rank's performance is  attributed to its first-stage training with a much larger dataset (1.5M queries). However, as shown in Table~\ref{table:trainingset}, the combination of both datasets leads to significant improvement, rendering RRK a highly competitive reranker. Interestingly, this combination hinders the convergence of the pointwise RRK. We suspect that the diversity in score distributions from different teachers is the complicating factor, whereas the listwise approach, leveraging ranking information, can effectively utilize multiple teachers.

\begin{table}[h]
    \centering
        \resizebox{0.5\textwidth}{!}{
   
    \begin{tabular}{lrr}
        \toprule
       \textbf{ Training set }&\textbf{Nb. queries} &\textbf{nDCG@10} \\
        \midrule
        MS-MARCO & 0.50M &57.7\\
        MS-MARCO & 0.15M &57.1\\
        E2RANK & 0.15M & 55.6\\
        MS MARCO + E2RANK &0.65M&58.4 \\
        \bottomrule
    \end{tabular}
    }
  
     \caption{RRK evaluation on BeIR (12 datasets) based on MS Marco and E2RANK \cite{E2rank2} datasets. The 0.15M MS-Marco subsamples produces better results than the E2Rank dataset while using a smaller teacher (0.6B/32B). Combining both improves by 0.7pt the nDCG@10 score.}
       \label{table:trainingset}
\end{table}

Our second ablation study focuses on how the PISCO compressor is generated. We obtain the best results by fine-tuning a pretrained PISCO model (compressor and decoder). Table~\ref{table:compressor} shows that the original \textbf{frozen} PISCO compressor yields poor results (55.5), whereas training the compressor from scratch jointly with the reranker (using a compressor and a reranker LORA adapter) leads to competitive performance (57.7). This demonstrates that RRK requires high-quality compressed representations to remain competitive.

\begin{table}[h]
    \centering 
        \resizebox{0.4\textwidth}{!}{
    \begin{tabular}{lr}
        \toprule
       \textbf{ Compressor configurations }&\textbf{nDCG@10 }\\
        \midrule
         Frozen PISCO compressor  &55.5\\
         Compressor from scratch&57.7\\
         fine-tuned PISCO compressor& 58.4\\
        \bottomrule
    \end{tabular}
    }
     \caption{Effectiveness of different PISCO compressor configurations. }
       \label{table:compressor}
\end{table}

\subsection{Comparison to PE-Rank and E2RANK}
\label{sec:perank}

The methods most closely related to RRK are PE-Rank \cite{liu2025leveraging} and E2RANK \cite{E2rank2}, both of which also aim to improve reranking efficiency through compressed document representations. Table~\ref{table:pee2r} compares RRK-MS and RRK with PE-Rank and E2RANK under their evaluation protocol: reranking the top 100 documents retrieved by BM25~\cite{robertson1996okapi}. We report results on the BeIR subsets used by both methods: TREC-COVID, SciFact, Web-Touché, NFCorpus, and DBPedia.
\begin{table}[h]
    \centering
        \resizebox{0.5\textwidth}{!}{
    \tiny
    \begin{tabular}{lrr}
        \toprule
        \textbf{Model} & \textbf{nDCG@10 }&\textbf{Lat. Ratio} \\
        \midrule
        RRK$\dagger$-MS & 55.4 & 1.0 (0.06)\\
        RRK$\dagger$ &56.5 & 1.0 (0.06)\\
        \midrule
       \textbf{E2RANK$\dagger$ (MS) }  & \\
        0.6B   & 53.9 & x2.1\\
        4B          & 56.2 & x7.0 \\
        8B          & 56.8 & x10.4 \\
        \midrule
        \textbf{E2RANK$\dagger$ (BGE)}\\
        0.6B & 55.0 & x2.1 \\
        4B           & 57.0 & x7.0 \\
        8B           & 57.2 & x10.4 \\
        \midrule
        PE-RANK$\dagger$ (7B,MS) & 51.3 & x7 \\
        \bottomrule
    \end{tabular}
    }

     \caption{RRK compared to E2RANK (trained resp. with MS-Marco and BGE datasets) and PE-RANK . Latency estimated from \cite{liu2025leveraging} and \cite{liu2025leveraging}}
         \label{table:pee2r}
\end{table}

\begin{table}[ht]
\centering
\resizebox{0.5\textwidth}{!}{
\begin{tabular}{lll}
    \toprule
    \textbf{Model} & \textbf{Input Length} & \textbf{Scoring Method} \\
    \midrule 
    RRK     & $2|q| + k(l+1)$        & cos($q$, $d_i$) \\
    PE-RANK & $|q| + k$              & $k$ decoding steps \\
    E2RANK  & $|q|+ 20\times|d| + k$ & cos($q$, $d_i$) \\
    \bottomrule
\end{tabular}}
\caption{Model Input Length and Scoring Methods for RRK, PE-RANK and E2RANK. $k$: number of reranked documents, $l$: number of PISCO memory tokens}
\label{tab:inputlength}
\end{table}
The approach most similar to RRK is PE-Rank. Although the two methods are conceptually related, they differ in both their foundations and their underlying assumptions. Following \citet{cheng2024xrag}, PE-Rank assumes that first-stage dense retrieval embeddings (Jina-Embeddings; \citealp{mohr2024multi}) already provide effective document representations for reranking. In other words, it treats compression inherited from an IR model as a suitable proxy. In contrast, RRK builds on the literature on soft compression for LLMs, which suggests that (a) representing documents with multiple tokens is beneficial, and (b) compression is most effective when learned within the LLM itself. RRK therefore assumes that compression learned through a question-answering objective yields more fine-grained representations than first-stage retrieval embeddings. Our experimental results support this hypothesis.

Regarding efficiency, one might expect PE-Rank to be faster, since it takes only $k$ tokens as input to rerank $k$ documents, whereas RRK uses 8 tokens per document. However, PE-Rank still relies on a final, slow decoding step to generate the ranked sequence of document identifiers, which prevents it from being faster than RRK (Table~\ref{tab:inputlength}).

E2RANK provides a different comparison point. It learns document embeddings shared by the first-stage retriever and the reranker. While this document representation is highly efficient, as it consists of a single embedding, its query representation is computationally expensive: to compute it, the top 20 documents (in text form) are concatenated with the query. As a result, the overall efficiency of the model is limited by the cost of processing roughly 20 documents. E2RANK overall good effectiveness compared to PE-Rank and RRK is certainly due to the use of E5 \cite{wang2024improving} a large dataset (1.5M queries) during its first-stage training.

\begin{figure}[h]
     \centering
      \includegraphics[width=\linewidth]{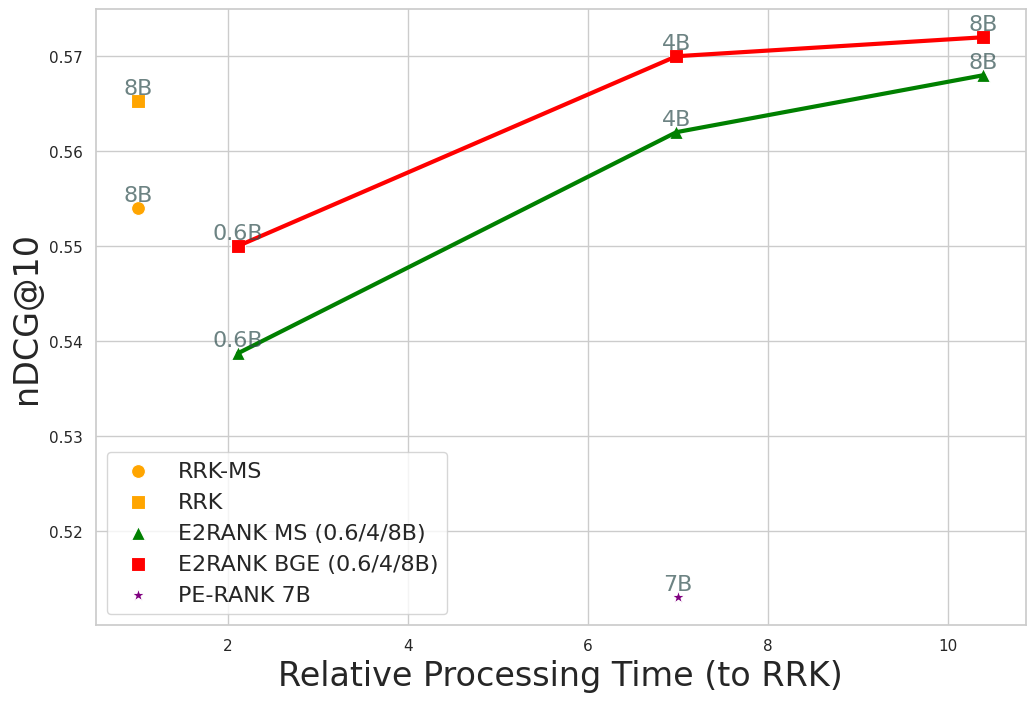}
     \caption{Efficiency/effectiveness comparison of RRK, PE-Rank, and E2RANK.}
     \label{fig:peE2RANK}
\end{figure}

Built on a richer compressed representation using 8 compression tokens, RRK consistently outperforms PE-Rank in both effectiveness and efficiency. It also surpasses E2RANK in speed while achieving comparable effectiveness on the shared datasets.


Overall, these comparisons highlight a key distinction between the approaches. Although all three methods rely on compressed document representations for efficient reranking, PE-Rank and E2RANK inherit compression from fixed IR embeddings, whereas RRK derives a soft, task-adaptive compression directly from the internal representations of a language model. This richer compressed representation preserves effectiveness without the degradation often observed with IR-based compression~\cite{cheng2024xrag,liu2025leveraging}.



\section{Conclusion}
In this work, we introduced RRK, a novel reranking framework that leverages compressed document representations to substantially improve efficiency while maintaining strong effectiveness. Our experiments show that using rich compressed embeddings based on a PISCO model—originally designed for Retrieval-Augmented Generation—enables RRK to achieve performance comparable to traditional, text-based rerankers, but with far lower latency, particularly on longer documents.

Importantly, our findings show that, although document compression can improve reranking efficiency, the quality of the compressed representation is critical. LLM-based compression yields fine-grained, expressive representations, while IR-based compression performs less effectively, likely because it loses important semantic information.

Overall, our results show that compressed representations enable an 8B-parameter model to run 3$\times$–18$\times$ faster than much smaller models (0.4–0.6B) while matching or outperforming their effectiveness.
These findings position LLM-based compression as a promising  approach for efficient reranking.

\section*{Limitations}

First, the efficiency of RRK is mostly due to its tiny input length. This advantage  holds  as long as the query itself is short. Using datasets like the BRIGHT dataset \cite{BRIGHT}, where queries length is comparable to BeIR document length,  breaks this advantage and makes the RRK model slow.

Secondly, it would be beneficial to employ smaller PISCO-based models instead of billion-sized ones as reranker. Unfortunately, our initial attempts to use smaller models, such as 1-4B parameter models, have not yet been successful. Using smaller models would lead to even better efficiency, and may reduce the index footprint (using smaller hidden dimensions).

Thirdly, the primary drawback of this method is the increased data storage requirement: each document requires \( c \times h \) storage, where \( c \) is the number of memory tokens and \( h \) is the hidden dimension of the model (3584 in our case). For the MS-MARCO collection (8.8 million documents), a PISCO model with 8 memory tokens results in a storage size of 230 GB when using \texttt{float16} encoding. It is about the size of the first version of Colbert indexes on MS-MARCO (286GB cf Table 4 in \citet{colbert}). We believe that additional optimizations such as quantization, are likely to reduce this footprint.

\bibliographystyle{acl_natbib}
\bibliography{latex/custom}

\newpage
\appendix
\onecolumn
\section{RRK Architecture}
\begin{figure*}[h]
    \centering
    \includegraphics[width=0.8\linewidth]{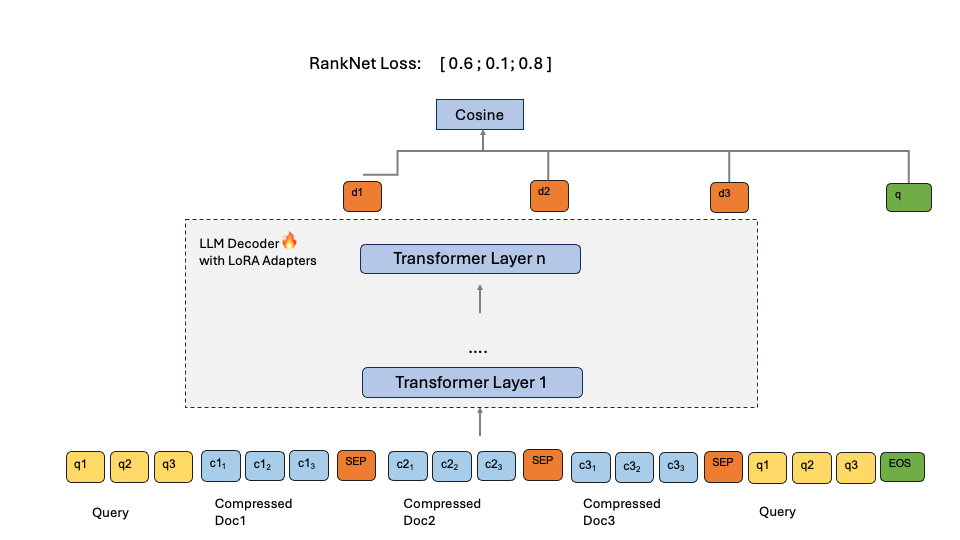}
    \caption{ RRK Architecture Schema}
    \label{fig:rrk-arch}
\end{figure*}

\section{Full comparison between RRK, PE-Rank and E2RANK models}

\begin{table*}[ht]
\centering

 \resizebox{0.85\textwidth}{!}{
\begin{tabular}{lcccccc}
    \toprule
    \textbf{Model} & \textbf{TREC-Covid} & \textbf{NFCorpus} & \textbf{Touché} & \textbf{DBPedia} & \textbf{SciFact} & \textbf{Avg} \\
    \midrule
    \textbf{E2 RANK (BGE)} \\
    0.6B & 79.2 & 38.6 & 41.9 & 42.0 & 73.4 & 55.0 \\
    4B   & 83.3  & 39.2 & \textbf{43.2} & 43.0 & 77.2 &\textbf{ 57.2} \\
    8B   & 84.1 & 39.1 & 42.2 & 43.4 & 77.5 &\textbf{ 57.2} \\
    \midrule
    \textbf{E2 RANK (MS)}\\
    0.6B & 80.0 & 37.6 & 36.6 & 41.9 & 73.2 & 53.9 \\
    4B   & 84.9  & 39.3 & 35.4 & 43.6 & \textbf{77.7 }& 56.2 \\
    8B   & 85.4 & \textbf{39.6} & 36.6 & 44.3 & 78.2 & 56.8 \\
    \midrule
    PE-RANK (MS)        & 77.5 & 36.4 & 33.1 & 40.1 & 69.4 & 51.3 \\
    \midrule
    RRK (MS)              & 81.5 & 37.9 &39.1 & 44.7 & 74.1 & 55.6 \\
    RRK (MS+BGE)          & \textbf{87.6} & 38.9 & 35.6 & \textbf{45.8} & 74.8 & 56.5 \\
    \bottomrule
\end{tabular}
}
\caption{RRK compared to E2RANK (trained resp. with MS-Marco and BGE datasets)  and PE-RANK.}
\end{table*}


\end{document}